\documentclass[12pt,a4paper,twoside]{article}
\newenvironment{changemargin}[2]{%
\begin{list}{}{%
\setlength{\leftmargin}{#1}%
\setlength{\rightmargin}{#2}%
}%
\item[]}
{\end{list}}
\usepackage{graphicx}
\usepackage{color}
\usepackage{lscape}
\usepackage{hyperref}
\usepackage{epstopdf}
\usepackage{lscape}
\usepackage{amsmath}
\usepackage{hyperref}
\usepackage{amsmath}
\usepackage{setspace}
\begin{document}
\baselineskip=0.30in
{\bf \LARGE

\begin{center}
\huge{Formula Method for Bound State Problems}
\end{center}
}
\vspace{4mm}
\begin{center}
{\Large{\bf B. J. Falaye $^a$$^{,}$$^\dag$$^{,}$}}\footnote{\scriptsize E-mail:~ fbjames11@physicist.net;~ babatunde.falaye@fulafia.edu.ng\\ \dag{Corresponding} author}\Large{\bf ,}  {\Large{\bf S. M. Ikhdair $^b$$^{,}$}}\footnote{\scriptsize E-mail:~ sameer.ikhdair@najah.edu;~ sikhdair@gmail.com.} \Large{\bf and} {\Large{\bf M. Hamzavi $^c$$^{,}$}}\footnote{\scriptsize E-mail:~ majid.hamzavi@gmail.com }
\end{center}
{\small
\begin{center}
{\it $^\textbf{a}$Applied Theoretical Physics Division, Department of Physics, Federal University Lafia,  P. M. B. 146, Lafia, Nigeria.}
{\it $^\textbf{b}$Department of Physics, Faculty of Science, an-Najah National University, New campus, P. O. Box 7, Nablus, West Bank, Palestine.}
{\it $^\textbf{c}$Department of Physics, University of Zanjan, Zanjan, Iran.}
\end{center}}

\begin{center}
Few-Body Syst, (2014) DOI 10.1007/s00601-014-0937-9
\end{center}

\begin{abstract}
\noindent
We present a simple formula for finding bound state solution of any quantum wave equation which can be simplified to the form of $\Psi''(s)+\frac{(k_1-k_2s)}{s(1-k_3s)}\Psi'(s)+\frac{(As^2+Bs+C)}{s^2(1-k_3s)^2}\Psi(s)=0$. The two cases where $k_3=0$ and $k_3\neq 0$ are studied. We derive an expression for the energy spectrum and the wave function in terms of generalized hypergeometric functions $_2F_1(\alpha, \beta; \gamma; k_3s)$. In order to show the accuracy of this proposed formula, we resort to obtaining bound state solutions for some existing eigenvalue problems in a rather more simplified way. This method has shown to be accurate, efficient, reliable and very easy to use particularly when applied to vast number of  quantum potential models.
\end{abstract}


{\bf PACs No.}:03.65.Ca; 03.65.Ge; 12.39.Fd; 03.65.Db

\begin{changemargin}{-1.2cm}{0.5cm}
\section{Introduction}
In quantum mechanics, while solving Schr\"{o}dinger, Dirac, Klein-Gordon, spinless Salpeter and Duffin-Kemmer-Petiau wave equations in the presence of some typical central or non central potential model, we do often come across differential equation of the form
\begin{equation}
\Psi''(s)+\frac{(k_1-k_2s)}{s(1-k_3s)}\Psi'(s)+\frac{(As^2+Bs+C)}{s^2(1-k_3s)^2}\Psi(s)=0,
\label{F1}
\end{equation}
after which an appropriate coordinate transformation of the form $s=s(r)$ has been used. The exact solutions of these wave equations with certain potentials have been of great interest. The analysis of such solutions makes the conceptual understanding of the quantum systems. The solutions provides a valuable means for checking and improving models and numerical methods introduced for solving complicated quantum systems. By solving time-independent Schr\"odinger equation for the energy eigenvalues and wave function, the structures and properties of atoms, molecules, solids, or any other steady-state forms of matter can be understood and explained. Such wave functions can describe the orbital motions of the electrons in the atoms. 

Up to now, several approaches have been developed for tackling equation of form (\ref{F1}). This include the asymptotic iteration method (AIM) [{1-25}], Feynman integral formalism [26-28], functional analysis method [24,29-31], exact quantization rule method [32-37], proper quantization rule [38-40], Nikiforov-Uvarov (NU) method [41-57], supersymetric quantum mechanics [24,58-63], the generalized pseudospectral method [64-69], etc. Notes on these techniques can be found in our recent work \cite{A1}.

\section{Statement of Problem}
The foregoing methods involve some complicated integrals while some involves great deal in algebraic manipulations. As a result, one needs a great efforts in mathematical skills before these methods could be applied to bound state problems. Thus, a simple efficient methodology will have its merit in Mathematical-Physics society, both for teaching and research purposes.
\section{Objectives}
\begin{itemize}
\item The objective of this research work is therefore to derive a very simple formula to exactly solve equation of form  (\ref{F1}) for energy eigenvalues and wave functions.  On contrary to other methods presented in the literature, this method is quite simple and could easily be followed, even by young researchers of Theoretical Physics solving bound state problems.

\item Furthermore, to test the accuracy of this new method, we investigate the solution of relativistic (Dirac and Klein-Gordon) and nonrelativistic Schr\"odinger equation for several central and non central potential models such as the harmonic oscillator, Manning-Rosen, Eckart, Kratzer-type, Coulomb, Morse and generalized non central Coulomb potentials for any arbitrary orbital quantum number $l \ne 0$.  It is worth mentioning that this method yields results which are in excellent agreement with the existing ones obtained by other approximate methods.
\end{itemize}
\section{Formula Method and their Applications}
We present a simple formula for finding bound state solutions of both the relativistic and nonrelativistic wave equations. The derivation of this method can be found in the appendix.  We also proceed to demonstrate the accuracy and convenience of this formula by finding energy eigenvalues and wave function of some already solved quantum systems, obtained by other methods early stated in the introductory text.
\subsection{Brief Overview of the Formula Method}
For a given Schr\"{o}dinger-like equation including the centrifugal barrier and/or the spin-orbit coupling term in the presence of any potential model which can be written in the form of equation (\ref{F1}), we propose that the energy eigenvalues and the corresponding wave function can be obtained by using the following formulas
\begin{subequations}
\begin{eqnarray}
&&\left[\frac{k_4^2-k_5^2-\left[\frac{1-2n}{2}-\frac{1}{2k_3}\left(k_2-\sqrt{(k_3-k_2)^2-4A}\right)\right]^2}{2\left[\frac{1-2n}{2}-\frac{1}{2k_3}\left(k_2-\sqrt{(k_3-k_2)^2-4A}\right)\right]}\right]^2-k_5^2=0, k_{3} \neq 0,
\label{F2A}\\
&&\Psi(s)=N_ns^{k_4}(1-k_3s)^{k_5}\ _2F_1\left(-n,\ n+2(k_4+k_5)+\frac{k_2}{k_3}-1;\ 2k_4+k_1,\ k_3s\right),
\label{F2B}
\end{eqnarray}
\end{subequations}
respectively, where
\begin{equation}
k_4=\frac{(1-k_1)+\sqrt{(1-k_1)^2-4C}}{2},\ \ 
k_5=\frac{1}{2}+\frac{k_1}{2}-\frac{k_2}{2k_3}+\sqrt{\left[\frac{1}{2}+\frac{k_1}{2}-\frac{k_2}{2k_3}\right]^2-\left[\frac{A}{k_3^2}+\frac{B}{k_3}+C\right]},
\label{F3}
\end{equation}
and $N_{n}$ is the normalization constant. Before proceeding to the applications of these formulas, let us first discuss a special case where $k_3\rightarrow0$. In this regard, the energy eigenvalues and the corresponding wave function can be obtained as
\begin{equation}
\left[\frac{B-k_4k_2-nk_2}{2k_4+k_1+2n}\right]^2-k_5^2=0,
\label{F4}
\end{equation}
and
\begin{equation}
\Psi(s)=N_ns^{k_4}exp(-k_5s)\ _1F_1\left(-n; 2k_4+k_1; (2k_5+k_2)s\right),
\label{F5}
\end{equation}
respectively.
\subsection{Applications}
In order to show the accuracy of these proposed formulas, we apply them to find the bound state solution of some quantum mechanical problems studied previously in the literature. Let us begin by obtaining the bound state solution of the Schr\"odinger equation for the following potential models. 
 
\subsubsection{Spherical Oscillator}
The spherical oscillator is an example for a problem with central field with a purely discrete spectrum. The Schr\"odinger equation with the spherically symmetric potential reads (\cite{BJ56} page 111)
\begin{equation}
-\frac{\hbar^2}{2m}\frac{1}{r^2}\frac{d}{dr}\left(r^2\frac{dR(r)}{dr}\right)+\left[\frac{\hbar^2\ell(\ell+1)}{2mr^2}+\frac{1}{2}m\omega^2r^2\right]R(r)=ER(r).
\label{F6}
\end{equation}
Now we recast the differential equation (\ref{F6}) into the form given in (\ref{F1}) by introducing the appropriate change of variables $r\rightarrow s$ through the mapping function $s=r^2$ to obtain
\begin{equation}
\frac{d^2R(s)}{ds^2}+\frac{3}{2s}\frac{dR(s)}{ds}+\left[\frac{\left(-\frac{m^2\omega^2}{4\hbar^2}\right)s^2+\left(\frac{mE}{2\hbar^2}\right)s-\frac{\ell}{4}(\ell+1)}{s^2}\right]R(s)=0.
\label{F7}
\end{equation}
Comparing equation (\ref{F7}) with equation (\ref{F1}), we obtain $k_1=3/2$, $k_2=k_3=0$, $A=-\frac{m^2\omega^2}{4\hbar^2}$, $B=\frac{mE}{2\hbar^2}$ and $C=-\frac{\ell}{4}(\ell+1)$.  Parameters $k_4$ and $k_5$ can be found as follows:
\begin{equation}
k_4=\frac{-\frac{1}{2}+\sqrt{\frac{1}{4}+\ell(\ell+1)}}{2}=\frac{\ell}{2},\ \ \ \ \ k_5=\sqrt{-A}=\sqrt{\frac{m^2\omega^2}{4\hbar^2}}=\frac{m\omega}{2\hbar}.
\label{F8}
\end{equation}
The eigenvalues of the Hamiltonian can now be expressed by using equation (\ref{F4}) as
\begin{equation}
\left[\frac{\frac{mE}{2\hbar^2}}{\ell+\frac{3}{2}+2n}\right]^2=\left(\frac{m\omega}{2\hbar}\right)^2\Leftrightarrow E_{n\ell}=\left(\ell+\frac{3}{2}+2n\right)\hbar\omega.
\label{F9}
\end{equation}
Using equation (\ref{F5}), the corresponding radial wave functions of arbitrary $\ell-$wave bound states of the Schr\"odinger equation with the spherical oscillator can be written as:
\begin{eqnarray}
R_{n\ell}(s)&=&N_{n\ell}s^{\frac{\ell}{2}}e^{-s\frac{m\omega}{2\hbar}}{_1F_1}\left(-n; \ell+\frac{3}{2}; \frac{m\omega}{\hbar}s\right)\ \ \ \mbox{or}\nonumber\\
R_{n\ell}(r)&=&N_{n\ell}r^{\ell}e^{-r^2\frac{m\omega}{2\hbar}}{_1F_1}\left(-n; \ell+\frac{3}{2}; \frac{m\omega}{\hbar}r^2\right),
\label{F10}
\end{eqnarray}
where $N_{n\ell}$ denotes the normalization factor. These results are in agreement with the ones obtained previously in ref. \cite{BJ56}.

\subsubsection{Manning-Rosen potential} 
The radial Schr\"odinger equation with the Manning-Rosen potential is given by \cite{BJ57}
\begin{equation}
\frac{d^2u_{n\ell}(r)}{dr^2}+\left[\frac{2\mu E_{n\ell}}{\hbar^2}-\frac{1}{b^2}\left(\frac{\alpha(\alpha-1)}{\left(e^{r/b}-1\right)^2}-\frac{\tilde{A}}{e^{r/b}-1}\right)-\frac{\ell(\ell+1)}{r^2}\right]u_{n\ell}(r)=0.
\label{F11}
\end{equation}
To solve the above equation for $\ell\neq0$ states, we apply the following approximation scheme \cite{BJ57}
\begin{equation}
\frac{1}{r^2}\approx\frac{1}{b^2}\left[D_0+D_1\frac{1}{e^{r/b}-1}+D_2\frac{1}{\left(e^{r/b}-1\right)^2}\right],
\label{F12}
\end{equation}
where $D_1=D_2=1$ and $D_0=1/12$ \cite{BJ57, BJ58, BJ59}. Now we recast the differential equation (\ref{F11}) into the form given in (\ref{F1}) by introducing the appropriate change of variables $r\rightarrow z$ through the mapping function $z=e^{-r/b}$ and defining
\begin{equation}
\xi_{n\ell}=\sqrt{-\frac{2\mu b^2E_{n\ell}}{\hbar^2}+\ell(\ell+1)D_0}, \ \ \ \ \ \beta_1=\tilde{A}-\ell(\ell+1)D_1\ \ \ \mbox{and} \ \ \ \beta_2=\alpha(\alpha-1)+\ell(\ell+1)D_2,
\label{F13}
\end{equation}
to obtain
\begin{equation}
\frac{d^2u_{n\ell}(z)}{dz^2}+\frac{(1-z)}{z(1-z)}\frac{du_{n\ell}(z)}{dz}+\left\{\frac{z^2(-\xi_{n\ell}^2-\beta_1-\beta_2)+z(2\xi_{n\ell}^2+\beta_1)+(-\xi_{n\ell}^2)}{z^2(1-z)^2}\right\}u_{n\ell}(z)=0.
\label{F14}
\end{equation}
Comparing equation (\ref{F14}) with equation (\ref{F1}), $k_1$, $k_2$, $k_3$, $A$, $B$ and $C$ can be easily determined together with $k_4$ and $k_5$ can be obtained as:
\begin{equation}
k_4=\xi_{n\ell}, \ \ \ \mbox{and}\ \ \ k_5=\frac{1}{2}(1+\sqrt{1+4\beta_2}).
\label{F15}
\end{equation}
Using equations (\ref{F2A}) and (\ref{F2B}), we can easily calculate the energy eigenvalues and the corresponding wave function as
\begin{equation}
E_{n\ell}=\frac{\hbar^2}{24\mu b^2}\ell(\ell+1)-\frac{\hbar^2}{2\mu b^2}\left[\frac{\tilde{A}+\alpha(\alpha-1)-\left[n+\frac{1}{2}+\frac{1}{2}\sqrt{(1-2\alpha)^2+4\ell(\ell+1)}\right]^2}{2\left[n+\frac{1}{2}+\frac{1}{2}\sqrt{(1-2\alpha)^2+4\ell(\ell+1)}\right]}\right]^2,
\label{F16}
\end{equation}
\begin{equation}
u_{n\ell}(z)=N_{n\ell}z^{\xi_{n\ell}}(1-z)^{\frac{1}{2}[1+\sqrt{1+4\beta_2}]}\ _2F_1\left(-n,n+2(\xi_{n\ell}+\frac{1}{2}[1+\sqrt{1+4\beta_2}]);2\xi_{n\ell}+1,z\right),
\label{F17}
\end{equation}
respectively, where $N_{n\ell}$ is the normalization constant. The above results are identical to the ones obtained in the literature \cite{BJ57, BJ60}. It should be noted that the solution of Hulth$\acute{e}$n potential can be easily obtained by setting $\alpha$ to 0 or 1 and $(A\hbar^2/2\mu b^2)$ to $Ze^2\delta$ in equations (\ref{F16}) and (\ref{F17}).

\subsubsection{Eckart potential} 
The radial Schr\"{o}dinger equation with the Eckart potential is given by \cite{BJ3, BJ61} (in units $\hbar=c=\mu=1$)
\begin{equation}
\frac{d^2R_{n\ell}(r)}{dr^2}+\left[2E_{n\ell}-\frac{2\beta e^{-r/a}}{\left(1-e^{-r/a}\right)^2}+\frac{2\alpha e^{-r/a}}{\left(1-e^{-r/a}\right)}-\frac{\ell(\ell+1)}{r^2}\right]R_{n\ell}(r)=0.
\label{F18}
\end{equation}
Using an appropriate approximation given in \cite{BJ3, BJ61} to deal with the centrifugal term and introducing a new variable of the form $z=e^{-r/a}$, we have
\begin{equation}
\frac{d^2R_{n\ell}(z)}{dz^2}+\frac{(1-z)}{z(1-z)}\frac{dR_{n\ell}(z)}{dz}+\left[\frac{z^2(2a^2[E_{n\ell}-\alpha])+z(2a^2[\beta-\alpha-2E_{n\ell}]-\ell(\ell+1))+2E_{n\ell}a^2}{z^2(1-z)^2}\right]R_{n\ell}(z)=0.
\label{F19}
\end{equation}
Now, after comparing equation (\ref{F16}) with equation (\ref{F1}), $k_1$, $k_2$, $k_3$, $A$, $B$ and $C$ can be easily determined and $k_4$ and $k_5$ can be obtained as:
\begin{equation}
k_4=\sqrt{-2E_{n\ell}a^2},\ \ \ \ \ k_5=\frac{1}{2}+\frac{1}{2}\sqrt{(2\ell+1)^2+8a^2\beta}.
\label{F20}
\end{equation}
Thus, the energy eigenvalues can be easily obtained by using formula (\ref{F2A}) as
\begin{equation}
E_{n\ell}=\alpha-\frac{1}{8a^2}\left[\frac{2\alpha a^2+\left[\left(n+\frac{1}{2}\right)+\frac{1}{2}\sqrt{(2\ell+1)^2+8\beta a^2}\right]^2}{\left[\left(n+\frac{1}{2}\right)+\frac{1}{2}\sqrt{(2\ell+1)^2+8\beta a^2}\right]}\right]^2,
\label{F21}
\end{equation}
and the corresponding wave functions can be obtained from (\ref{F2B}) as:
\begin{eqnarray}
&&R_{n\ell}(z)=N_{n\ell}z^{\sqrt{-2E_{n\ell}a^2}}(1-z)^{\frac{1}{2}+\frac{1}{2}\sqrt{(2\ell+1)^2+8a^2\beta}}\nonumber\\
&&\times_2F_1\left(-n, n+2\left(\sqrt{-2E_{n\ell}a^2}+\frac{1}{2}+\frac{1}{2}\sqrt{(2\ell+1)^2+8a^2\beta}; 2\sqrt{-2E_{n\ell}a^2}+1,z\right)\right),
\label{F22}
\end{eqnarray}
where $N_{n\ell}$ is the normalization constant. These results are in excellent agreement with the ones obtained previously \cite{BJ3, BJ61}
\subsubsection{Kratzer potential} 
The radial Schr\"{o}dinger equation with the Kratzer potential is given by \cite{BJ7} 
\begin{equation}
\frac{d^2R_{n\ell}(s)}{ds^2}+\left[\frac{s^2\left(\frac{2\mu E_{n\ell}}{\hbar^2}\right)+s\left(\frac{4\mu D_{e}a}{\hbar^2}\right)+\left(-\frac{2\mu D_ea^2}{\hbar^2}-\ell(\ell+1)\right)}{s^2}\right]R_{n\ell}(s)=0,
\label{F23}
\end{equation}
where we have introduced a new transformation of the form $r=s$. Now by comparing equation (\ref{F23}) with equation (\ref{F1}), we can easily find the values of parameters $A$, $B$ and $C$. It is also clear that $k_1=k_2=k_3=0$ and the other parameters are obtained as
\begin{equation}
k_4=\frac{1}{2}+\sqrt{\left(\frac{1}{2}+\ell\right)^2+\frac{2\mu D_ea^2}{\hbar^2}}\ \ \mbox{and}\ \ \ k_5=\sqrt{-\frac{2\mu E_{n\ell}}{\hbar^2}}.
\label{F24}
\end{equation}
The energy eigenvalues can now be obtained by means of equation (\ref{F4}) as
\begin{equation}
E_{n\ell}=-\frac{2\mu D_e^2a^2}{\hbar^2}\left[n+\frac{1}{2}+\sqrt{\left(\frac{1}{2}+\ell\right)^2+\frac{2\mu D_ea^2}{\hbar^2}}\right]^{-2},
\label{F25}
\end{equation}
and the corresponding wave function can be found through equation (\ref{F5}) as:
\begin{eqnarray}
R_{n\ell}(r)=r^{\frac{1}{2}+\sqrt{\left(\frac{1}{2}+\ell\right)^2+\frac{2\mu D_ea^2}{\hbar^2}}}exp\left(-\sqrt{-\frac{2\mu E_{n\ell}}{\hbar^2}}\right)\nonumber\\ \times _1F_1\left(-n; 1+2\left(\sqrt{\left(\frac{1}{2}+\ell\right)^2+\frac{2\mu D_ea^2}{\hbar^2}}\right);2r\sqrt{-\frac{2\mu E_{n\ell}}{\hbar^2}}\right).
\label{F26}
\end{eqnarray}
These results are identical with the ones obtained in refs \cite{BJ7,BJ62}.

\subsubsection{Generalized non central Coulomb potential model}
The ring-shaped coulombic (Hartmann) potential in spherical coordinates is given by \cite{BJ63, BJ64, BJ65}
\begin{equation}
V(r,\theta)=-\frac{Ze^2}{r}+\frac{\beta}{r^2\sin^2\theta}+\frac{\gamma\cos\theta}{r^2\sin^2\theta}.
\label{F27}
\end{equation}
The solution of the Schr\"{o}dinger equation with the combined Coulomb plus Aharanov-Bohm potential and the Hartmann ring-shaped potential, which was originally proposed as a model for benzene molecules, can be obtained as a special case of potential (\ref{F27}). Now, the Schr\"{o}dinger equation in the presence of potential $V(r,\theta)$ can be reduced to the two ordinary differential equations \cite{BJ63, BJ64, BJ65}
\begin{subequations}
\begin{eqnarray}
\frac{d^2R_{n\ell}(r)}{dr^2}+\frac{2}{r}\frac{dR_{n\ell}(r)}{dr}+\left[\frac{2\mu}{\hbar^2}\left(E_{n\ell}+\frac{Ze^2}{r}\right)-\frac{\ell(\ell+1)}{r^2}\right]R_{n\ell}(r)&=&0,
\label{F28A}\\
\frac{d^2\Theta(\theta)}{d\theta^2}+\coth\theta\frac{d\Theta(\theta)}{d\theta}+\left[\ell(\ell+1)-\left(\frac{m^2+\beta+\gamma\cos\theta}{\sin^2\theta}\right)\right]\Theta(\theta)&=&0,
\label{F28B}
\end{eqnarray}
\end{subequations}
where the corresponding total wave function is taken as $\Psi(r,\theta,\phi)=R(r)\Theta(\theta)e^{im\varphi}$. Now by introducing a new transformation of the form $s=r$ and $s=\frac{\cos\theta-1}{2}$ in equations (\ref{F28A}) and (\ref{F28B}) respectively, we obtain 
\begin{subequations}
\begin{equation}
\frac{d^2R_{n\ell}(s)}{dr^2}+\frac{2}{s}\frac{dR_{n\ell}(s)}{dr}+\left[\frac{\left(\frac{2\mu E}{\hbar^2}\right)s^2+\left(\frac{2\mu Ze^2}{\hbar^2}\right)s+(-\ell(\ell+1))}{s^2}\right]R_{n\ell}(s)=0,
\label{F29A}
\end{equation}
\begin{equation}
\frac{d^2\Theta(s)}{ds^2}+\frac{2s+1}{s(1+s)}\frac{d\Theta(s)}{ds}+\left[\frac{(-\ell(\ell+1))s^2+(-\ell(\ell+1)-\gamma/2)s+(-[m^2+\beta+\gamma]/4)}{s^2(1+s)^2}\right]\Theta(s)=0.
\label{F29B}
\end{equation}
\end{subequations}
It can be deduced from equation (\ref{F29A}) that $k_1=2, k_2=k_3=0$ and the values of $A$, $B$, $C$ can be clearly seen. The parameters $k_4$ and $k_5$ can therefore be calculated as
\begin{equation}
k_4=\ell \ \ , \ \ k_5=\sqrt{-\frac{2\mu E}{\hbar^2}},
\label{F30}
\end{equation}
and hence the eigenvalues equation can be found by using formula (\ref{F4}) as
\begin{equation}
E_{N\ell}=-\frac{\mu Z^2e^4}{2\hbar^2[N+\ell+1]^2}.
\label{F31}
\end{equation}
Taking $N+\ell+1=n$, this result is in agreement with the ones obtained for hydrogen atom in ref. (\cite{BJ56} page 118)  when $Z=1$ and ref. (\cite{BJ66} page 133) when $Z=1/4\pi\epsilon_0$. Furthermore, by comparing equation (\ref{F29B}) with equation (\ref{F1}), $k_1$, $k_2$, $k_3$, $A$, $B$ and $C$ can be easily determined then $k_4$ and $k_5$ can be obtained as:
\begin{equation}
k_4=\frac{1}{2}\sqrt{m^2+\beta+\gamma},\ \ \ \ \ k_5=\frac{1}{2}\sqrt{m^2+\beta-\gamma}, 
\label{F32}
\end{equation}
and by using equation (\ref{F2A}) the following results can be obtained:
\begin{eqnarray}
(\ell-n)^2&=&\left[\frac{1}{2}\sqrt{m^2+\beta+\gamma}+\frac{1}{2}\sqrt{m^2+\beta-\gamma}\right]^2\nonumber\\
\ell&=&n+\left[\frac{(m^2+\beta)+\sqrt{(m^2+\beta)^2-\gamma^2}}{2}\right]^{1/2}.
\label{F33}
\end{eqnarray}
Making use of equations (\ref{F33}) and (\ref{F31}), the final energy levels for a real bound charged particle in a Coulombic field plus a combination of non central potentials given by equation (\ref{F28B}) are
\begin{equation}
E_{n,N,m}=-\frac{\mu Z^2e^4}{2\hbar^2\left[N+1+n+\left[\frac{(m^2+\beta)+\sqrt{(m^2+\beta)^2-\gamma^2}}{2}\right]^{1/2}\right]^2},
\label{F34}
\end{equation}
and the complete eigenfunctions (radial$\times$angular) can be obtained from formulas (\ref{F2B} and \ref{F5}) as
\begin{eqnarray}
\Psi(r,\theta,\phi)&=&N_{n\ell}r^\ell\exp\left(-\sqrt{\frac{-2\mu E}{\hbar^2}}\right)\left(\frac{\cos\theta-1}{2}\right)^{\frac{1}{2}\sqrt{m^2+\beta+\gamma}}\left(\frac{\cos\theta+1}{2}\right)^{\frac{1}{2}\sqrt{m^2+\beta-\gamma}}e^{\pm im\varphi}\nonumber\\
&&\times\ _2F_1\left(-n,n+\sqrt{m^2+\beta+\gamma}+\sqrt{m^2+\beta-\gamma}+1;1+\sqrt{m^2+\beta+\gamma},\left(\frac{1-\cos\theta}{2}\right)\right)\nonumber\\
&&\times\ _1F_1\left(-n; 2(\ell+1); 2r\sqrt{\frac{-2\mu E}{\hbar^2}}\right)
\label{F35}.
\end{eqnarray}
These results are in agreement with the ones obtained previously \cite{BJ63,BJ64,BJ65, BJ67}. To show the effectiveness and flexibility of our approach, we extend our applications to solve the Klein-Gordon equation for the following potential models.

%%%%%%%%%%%%%%%%%%%%%%%%%%%%%%%%%%%%%%%%%%%%%%%%%%%%%%%%%%%%%%%%%%%%%%%%%%%%%%%%%%%%%%%%%%%%%%%%%%%%%%%%%%%%%%%%%%%%%%%%%%%%%%%
%%%%%%%%%%%%%%%%%%%%%%%%%%%%%%%%%%%%%%%%%%%%%%%%%%%%%%%%%%%%%%%%%%%%%%%%%%%%%%%%%%%%%%%%%%%%%%%%%%%%%%%%%%%%%%%%%%%%%%%%%%%%%%%
\subsubsection{Coulomb potential}
The Klein-Gordon equation with the Coulomb potential is given by equation (1) in page 45 of ref. \cite{BJ68}
\begin{equation}
\left[\frac{d^2}{dr^2}-\frac{\ell(\ell+1)-(Z\alpha)^2}{r^2}+\frac{2\epsilon Z\alpha}{\hbar cr}-\frac{m_0^2c^4-\epsilon^2}{\hbar^2c^2}\right]R_\ell(r)=0.
\label{F36}
\end{equation}
Now, let us recast the differential equation (\ref{F36}) into the form given in equation (\ref{F1}) to obtain
\begin{equation}
\frac{d^2R_\ell(r)}{dr^2}+\left[\frac{(Z\alpha)^2-\ell(\ell+1)+\left(\frac{2\epsilon Z\alpha}{\hbar c}\right)r+\left(\frac{\epsilon^2-m_0^2c^4}{\hbar^2c^2}\right)r^2}{r^2}\right]R_\ell(r)=0.
\label{F37}
\end{equation}
Comparing equation (\ref{F37}) with equation (\ref{F1}), we obtain $k_1=k_2=k_3=0$, $A=(Z\alpha)^2-\ell(\ell+1)$, $B=\frac{2\epsilon Z\alpha}{\hbar c}$ and $C=\frac{\epsilon^2-m_0^2c^4}{\hbar^2c^2}$. Thus, parameters $k_4$ and $k_5$ can be easily found as follows:
\begin{equation}
k_4=\frac{1}{2}+\sqrt{\left(\ell+\frac{1}{2}\right)-(Z\alpha)^2}=\frac{1}{2}+\mu,\ \ \ \ \ k_5=\sqrt{\frac{m_0^2c^4-\epsilon^2}{\hbar^2c^2}}=\sigma,
\label{F38}
\end{equation}
where $\mu=\sqrt{\left(\ell+\frac{1}{2}\right)-(Z\alpha)^2}$. The relativistic energy eigenvalues can now be expressed by using equation (\ref{F4}) as
\begin{equation}
\left[\frac{\frac{2\epsilon Z\alpha}{\hbar c}}{1+2\sqrt{\left(\ell+\frac{1}{2}\right)-(Z\alpha)^2}+2n}\right]^2=\frac{m_0^2c^4-\epsilon^2}{\hbar^2c^2}\Leftrightarrow \frac{\epsilon Z\alpha}{m_0^2c^4-\epsilon^2}=n'+\frac{1}{2}+\sqrt{\left(\ell+\frac{1}{2}\right)-(Z\alpha)^2},
\label{F39}
\end{equation}
or more explicitly,
\begin{equation}
\epsilon_{n'\ell}=-m_0c^2\left[1+\frac{(Z\alpha)^2}{\left(n'+\frac{1}{2}+\left[\left(\ell+\frac{1}{2}\right)^2-(Z\alpha)^2\right]^{1/2}\right)^2}\right]^{-\frac{1}{2}}
\label{F40}
\end{equation}
Using equation (\ref{F5}), the correspondingly radial wave functions of arbitrary $\ell-$wave bound states of the Klein-Gordon equation with the Coulomb potential can be written as:
\begin{eqnarray}
R_\ell(r)=N_{n\ell}r^{\mu+\frac{1}{2}}e^{-\sigma}{_1F_1}\left(-n; 2\mu+\frac{1}{2}; 2\sigma r\right)
\label{F41}
\end{eqnarray}
These results are in agreement with the ones obtained previously in ref. \cite{BJ68}.

\subsubsection{Eckart potential}
The Klein-Gordon equation with the Eckart potential is given by equation (8) of ref \cite{BJ69}
\begin{equation}
x^2\frac{d^2u(x)}{dx^2}+x\frac{du(x)}{dx}-\left[\lambda^2-\frac{k^2\alpha x}{1-x}+\frac{k^2\beta x}{(1-x)^2}+\frac{\ell(\ell+1)x}{(1-x)^2}\right]u(x)=0,
\label{F42}
\end{equation}
where variable $x$ is defined as $x = e^{-r/a}$, $\lambda^2=a^2(M^2-E^2)$, $k^2=2(M+E)a^2$. Now, let us simplify equation (\ref{F42}) into the form of equation (\ref{F1})
\begin{equation}
\frac{d^2u(x)}{dx^2}+\frac{1}{x}\frac{du(x)}{dx}+\left[\frac{x^2[-\lambda^2-k^2\alpha]+x\left[2\lambda^2+k^2(\alpha-\beta)-\ell(\ell+1)\right]-\lambda^2}{x^2(1-x^2)}\right]u(x)=0.
\label{F43}
\end{equation}
Comparing equation (\ref{F43}) with equation (\ref{F1}), we obtain $k_1=1$, $k_2=1$, $k_3=1$, $A=-\lambda^2-k^2\alpha$, $B=2\lambda^2+k^2(\alpha-\beta)-\ell(\ell+1)$ and $C=-\lambda^2$. Thus, parameters $k_4$ and $k_5$ can be easily obtained as follows:
\begin{equation}
k_4=\lambda=\sqrt{a^2(M^2-E^2)},\ \ \ \ \ k_5=\frac{1}{2}+\frac{1}{2}\sqrt{(2\ell+1)^2+4k^2\beta}=\frac{1}{2}+\frac{1}{2}\sqrt{(2\ell+1)^2+8a^2(M+E)\beta}=\frac{1}{2}+\delta, 
\label{F44}
\end{equation}
where $\delta=\frac{1}{2}\sqrt{(2\ell+1)^2+8a^2(M+E)\beta}$. The relativistic energy eigenvalues can now be expressed by using equation (\ref{F2A}) or (\ref{A15}) as
\begin{equation}
a\sqrt{(M^2-E^2)}+\sqrt{\left(\ell+\frac{1}{2}\right)^2+2a^2\beta(M+E)}-a\sqrt{M^2-E^2+2\alpha(M+E)}+\frac{1}{2}+n=0.
\label{F45}
\end{equation}
Using equation (\ref{F2B}), the correspondingly radial wave functions of arbitrary $\ell-$wave bound states of the Klein-Gordon equation with the mixed Eckart potentials can be written as:
\begin{eqnarray}
u_{n\ell}(r)=N_{n\ell}e^{-\lambda r/a}\left(1-e^{-r/a}\right)^{\delta+1}{_2F_1}\left(-n, n+2\left(\lambda+\frac{1}{2}+\delta\right); 2\lambda+1, e^{-r/a}\right).
\label{F46}
\end{eqnarray}
These results are in agreement with the ones obtained previously in ref. \cite{BJ69}. Finally we solve the Dirac equation for the Morse potential and also Kemmer Equation for Dirac Oscillator.
%%%%%%%%%%%%%%%%%%%%%%%%%%%%%%%%%%%%%%%%%%%%%%%%%%%%%%%%%%%%%%%%%%%%%%%%%%%%%%%%%%%%%%%%%%%%%%%%%%%%%%%%%%%%%%%%%
%%%%%%%%%%%%%%%%%%%%%%%%%%%%%%%%%%%%%%%%%%%%%%%%%%%%%%%%%%%%%%%%%%%%%%%%%%%%%%%%%%%%%%%%%%%%%%%%%%%%%%%%%%%%%%
\subsubsection{Morse potential}
The Dirac equation with the Morse potential is given by equation (33) of ref \cite{BJ70}
\begin{equation}
\frac{d^2G_{n\kappa}(y)}{dy^2}+\frac{1}{y}\frac{dG_{n\kappa}(y)}{dy}+\left[-\frac{\epsilon^2}{y^2}+\frac{\beta_1^2}{y}-\beta_2^2\right]G_{n\kappa}(y)=0,
\label{F47}
\end{equation}
where the parameters $y$, $\epsilon$, $\beta_1$ and $\beta_2$ have been defined in ref. \cite{BJ70}.  Now we recast the differential equation (\ref{F47}) into the form given in (\ref{F1})  to obtain
\begin{equation}
\frac{d^2G_{n\kappa}(y)}{dy^2}+\frac{1}{y}\frac{dG_{n\kappa}(y)}{dy}+\left[\frac{-\epsilon^2+\beta_1^2y-\beta_2y^2}{y^2}\right]G_{n\kappa}(y)=0.
\label{F48}
\end{equation}
By comparing equation (\ref{F48}) with equation (\ref{F1}), we obtain $k_1=1$, $k_2=k_3=0$, $A=-\beta_2^2$, $B=\beta_1^2$ and $C=-\epsilon^2$.  Parameters $k_4$ and $k_5$ can be found as follows:
\begin{equation}
k_4=\sqrt{-C}=\epsilon,\ \ \ \ \ k_5=\sqrt{-A}=\beta_2.
\label{F49}
\end{equation}
The eigenvalues of the Hamiltonian can now be expressed by using equation (\ref{F4}) as
\begin{equation}
\left[\frac{\beta_1^2}{2\epsilon+1+n}\right]^2=\beta_2^2\Leftrightarrow \epsilon_{n}=\frac{\beta_1^2-(2n+1)\beta_2}{2\beta_2}.
\label{F50}
\end{equation}
Using equation (\ref{F5}), the corresponding radial wave functions for the Dirac equation with the Morse potential can be written as:
\begin{eqnarray}
R_{n\ell}(r)&=&N_{n\kappa} y^{\epsilon}e^{-\beta_2y}{_1F_1}\left(-n; 2\epsilon+1; 2\beta_2y\right),
\label{F51}
\end{eqnarray}
where $N_{n\kappa}$ denotes the normalization constant. These results are in agreement with the ones obtained previously in ref. \cite{BJ70}.

%%%%%%%%%%%%%%%%%%%%%%%%%%%%%%%%%%%%%%%%%%%%%%%%%%%%%%%%%%%%%%%%%%%%%%%%%%%%%%%%%%%%%%%%%%%%%%%%%%%%%%%%%%%%%%
\subsubsection{Kemmer Equation for Dirac Oscillator}
The relativistic Kemmer equation for Dirac oscillator is given as \cite{BJ71}
\begin{equation}
\left[\frac{d^2}{dr^2}-\frac{M^2\omega^2r^2}{\hbar^2}-\frac{\ell(\ell+1)}{r^2}+\frac{E^2-\left(Mc^2/2\right)}{\hbar^2c^2}-\frac{M\omega}{\hbar}\left[j(j+1)-\ell(\ell+1)+1\right]\right]R_{n\ell}(r)=0.
\label{F52}
\end{equation}
Now we transform the differential equation (\ref{F52}) into the form given in (\ref{F1}) by introducing the appropriate change of variables $r\rightarrow \xi$ through the mapping function $\xi=\sqrt{M\omega/\hbar}r$ to obtain
\begin{equation}
\left[\frac{d^2}{d\xi^2}+\frac{1}{2\xi}\frac{d}{d\xi}+\frac{\left(-\frac{1}{4}\right)\xi^2+\frac{\varsigma}{2}\xi-\frac{1}{4}\ell(\ell+1)}{\xi^2}\right]R_{n\ell}(\xi)=0,
\label{F53}
\end{equation}
where $\varsigma=\frac{E^2-\left(Mc^2/2\right)}{2\hbar\omega Mc^2}-\frac{1}{2}\left[j(j+1)-\ell(\ell+1)+1\right]$ has been introduced for simplicity. By comparing equation (\ref{F53}) with equation (\ref{F1}), we obtain $k_1=1/2$, $k_2=k_3=0$, $A=-\frac{1}{4}$, $B=\frac{\varsigma}{2}$ and $C=-\frac{1}{4}\ell(\ell+1)$.  Parameters $k_4$ and $k_5$ can be found as follows:
\begin{equation}
k_4=\frac{1}{2}(\ell+1),\ \ \ \ \ k_5=\sqrt{-A}=\frac{1}{2}.
\label{F54}
\end{equation}
The eigenvalues of the Hamiltonian can now be expressed by using equation (\ref{F4}) as
\begin{equation}
\left[\frac{\frac{\varsigma}{2}}{\ell+\frac{3}{2}+2n}\right]^2=\frac{1}{4} \Leftrightarrow \varsigma_{n}=\ell+\frac{3}{2}+2n,
\label{F55}
\end{equation}
or more explicitly as
\begin{equation}
E=\frac{1}{2}Mc^2\left\{1+4\left[4(n+1)+j(j+1)-\ell(\ell-1)\right]\frac{\hbar\omega}{Mc^2}\right\}^{\frac{1}{2}}.
\label{F56}
\end{equation}
By using equation (\ref{F5}), the correspondingly radial wave functions  can be written as:
\begin{eqnarray}
R_{n\ell}(r)&=&N_{n\ell}\xi^{\frac{1}{2}(\ell+1)}e^{-\frac{\xi}{2}}{_1F_1}\left(-n; \ell+\frac{3}{2}; \xi\right),
\label{F57}
\end{eqnarray}
where $N_{n\ell}$ denotes the normalization constant. These results are in agreement with the ones obtained previously in ref. \cite{BJ71}.

%%%%%%%%%%%%%%%%%%%%%%%%%%%%%%%%%%%%%%%%%%%%%%%%%%%%%%%%%%%%%%%%%%%%%%%%%%%%%%%%%%%%%%%%%%%

\section{Concluding Remarks}
We proposed a simple formula for finding eigensolutions (energy eigenvalues and wave function) of any non-relativistic and relativistic wave equations. The proposed formula is derived via the asymptotic iteration method (AIM) and functional analysis approach (FAA). This approach presents a new alternative and accurate method to obtain bound states with different potential models. To show the accuracy and effectiveness of this method, we re-obtained the bound state solution of non-relativistic and relativistic wave equations with a vast number of central and non-central potential models like harmonic oscillator, Manning-Rosen,  Hulth$\acute{e}$n, Eckart, Kratzer-type, Coulomb, Morse and generalized non central Coulomb potentials. We considered the cases of $k_3=0$ and $k_3\ne 0$ for exact and approximate solutions. It is worth being paid attention that all of  our results are in excellent agreement with the ones obtained previously by other methods. 

This method is recommended for both teaching and research purposes.

\section*{Acknowledgments}
We thank the kind referee for the positive enlightening comments and suggestions, which have greatly helped us in making improvements to this paper. In addition, BJF acknowledges eJDS (ICTP).

\section*{Appendix A: Derivation of the Formula Method}
Firstly, let us analyze the asymptotic behavior at the origin and at infinity for the finiteness of our solution. It can be tested when $s\rightarrow0$ by taking the solution of equation (\ref{F1}) as $\Psi(s)=s^{k_4}$, where
\begin{equation}
k_4=\frac{(1-k_1)+\sqrt{(1-k_1)^2-4C}}{2}.\tag{A1}
\label{A1}
\end{equation}
Again, it can also be proved that when $s\rightarrow\frac{1}{k_3}$, the solution of equation (\ref{F1}) is $\Psi(s)=(1-k_3s)^{k_5}$, where
\begin{equation}
k_5=\frac{1}{2}+\frac{k_1}{2}-\frac{k_2}{2k_3}+\sqrt{\left[\frac{1}{2}+\frac{k_1}{2}-\frac{k_2}{2k_3}\right]^2-\left[\frac{A}{k_3^2}+\frac{B}{k_3}+C\right]}.\tag{A2}
\label{A2}
\end{equation}
Hence, the wave function in the intermediate region, for this problem, can be taken as
\begin{equation}
\Psi(s)=s^{k_4}(1-k_3s)^{k_5}F(s).\tag{A3}
\label{A3}
\end{equation}
The newly defined function $F(s)$ should turn out to be something very simple such as, presumably, polynomials in $s$. It is worthwhile to make a last substitution aiming at converting the differential equation for $\Psi(s)$ to one for $F(s)$. The substitution of the above equation (\ref{A3}) into equation (\ref{A1}) leads to the following second-order differential equations 
\begin{subequations}
\begin{align*}
&&F''(s)+F'(s)\left[\frac{(2k_4+k_1)-sk_3(2k_4+2k_5+\frac{k_2}{k_3})}{s(1-k_3s)}\right]\nonumber\\
&&-\left[\frac{2k_3k_4(k_4-1)+k_5k_3(2k_4+k_1)+k_4(k_1k_3+k_2)-B}{s(1-k_3s)}\right]F(s)=0,\tag{A4a}
\label{A4A}
\end{align*}
\begin{equation}
{\footnotesize
F''(s)+F'(s)\left[\frac{(2k_4+k_1)-sk_3(2k_4+2k_5+\frac{k_2}{k_3})}{s(1-k_3s)}\right]-\left[\frac{k_3(k_4+k_5)^2+(k_4+k_5)(k_2-k_3)+\frac{A}{k_3}}{s(1-k_3s)}\right]F(s)=0.\tag{A4b}
\label{A4B}}
\end{equation}
\end{subequations}
It should be noted that equation (\ref{A4A}) is equivalent to equation (\ref{A4B}), but equation (\ref{A4A}) becomes more complicated during the course of our calculations. Therefore, we shall continue our derivation with equation (\ref{A4B}). At this stage, we shall obtain the solution of equation (\ref{A4B}) by using the asymptotic iteration method and functional analysis method
%%%%%%%%%%%%%%%%%%%%%%%%%%%%%%%%%%%%%%%%%%%%%%%%%%%%%%%%%%%%%%%%%%%%%%%%%%%%%%%%%%%%%%%%%%%%%%%%%%%%%%
\subsection*{Derivation via AIM}
The asymptotic iteration method (AIM) for eigenvalues problem has been introduced by Ciftci et al. \cite{BJ1,BJ2}. Ever since then, it has been used in many physical systems to obtain the whole spectra (\cite{BJ24} and refs. therein). The beauty about the method is that it reproduces exact solutions of many exactly solvable quantum systems and also gives accurate result for the non solvable potentials such as sextic oscillator, deformed Coulomb potential, etc.

Because of unfamiliar readers, firstly, we give a brief review of AIM with all necessary formulas for our derivation. For a given potential the idea is to convert the Schr\"{o}dinger-like equation to the homogenous linear second-order differential equation of the form:
\begin{equation}
y_n''(x)=\lambda_0(x)y_n'(x)+s_0(x)y_n(x),\tag{A5}
\label{A5}
\end{equation}
where $\lambda_0(x)\neq0$ and the prime denote the derivative with respect to $x$, the extral parameter $n$ denotes the radial quantum number. The variables, $s_0(x)$ and  $\lambda_0(x)$ are sufficiently differentiable. To find energy spectrum equation of any Schrodinger-like equation, the equation is first transformed to form of (\ref{A5}). Then, one need to obtain $\lambda_k(x)$ and $s_k(x)$ with $k>0$, i.e.,
\begin{equation}
\lambda_k(x)=\lambda_{k-1}'(x)+s_{k-1}(x)+\lambda_o(x)\lambda_{k-1}(x),\ \ \ \ 
s_k(x)=s_{k-1}'(x)+s_o(x)\lambda_{k-1}(x).\tag{A2}
\label{A6}
\end{equation}
With $\lambda_k(x)$ and $s_k(x)$ values, one can obtain the quantization condition
\begin{equation}\tag{A7}
\delta_k(x)=
\left|
\begin{array}{lr}     
\lambda_k(x)&s_k(x) \\      
  \lambda_{k-1}(x)&s_{k-1}(x)
  \end{array}
  \right|=0\ \ ,\ \  \ k=1, 2, 3....
\label{A7} 
  \end{equation}
The energy eigenvalues are then obtained by the condition given by equation (\ref{A7}) if the problem is exactly solvable. For nontrivial potentials that have no exact solutions, for a specific n principal quantum number, we choose a suitable $x_0$ point, determined generally as the maximum value of the asymptotic wave function or the minimum value of the potential and the approximate energy eigenvalues are obtained from the roots of equation (\ref{A7}) for sufficiently great values of $k$ with iteration for which
$k$ is always greater than $n$ in the numerical solutions. Furthermore, the eigenfunction can be obtained by transforming the Schr\"{o}dinger-like equation to the form of
\begin{equation}
y''(x)=2\left(\frac{ax^{N+1}}{1-bx^{N+2}}-\frac{m+1}{x}\right)y'(x)-\frac{Wx^N}{1-bx^{N+2}}y(x). \tag{A8}
\label{A8}
\end{equation}
The exact solution $y_n(x)$ can be expressed as \cite{BJ1,BJ2}
\begin{equation}
y_n(x)=(-1)^nC_2(N+2)^n(\sigma)_{_n}{_2F_1(-n,\rho+n;\sigma;bx^{N+2})}, \tag{A9}
\label{A9}
\end{equation}
where the following notations has been used
\begin{equation}
(\sigma)_{_n}=\frac{\Gamma{(\sigma+n)}}{\Gamma{(\sigma)}}\ \ ,\ \ \sigma=\frac{2m+N+3}{N+2}\ \ \mbox{and}\ \ \rho=\frac{(2m+1)b+2a}{(N+2)b}. \tag{A10}
\label{A10}
\end{equation}
Now, comparing equation (\ref{A4B}) with equation (\ref{A9}), we can determine the following parameters
\begin{equation}
m=k_4+\frac{k_1}{2}-1,\ \ \ \ a=\left(k_5-\frac{k_1}{2}+\frac{k_2}{2k_3}\right)k_3,\ \ \ \ \sigma=2k_4+k_1\ \ \mbox{and}\ \ \rho=2(k_4+k_5)+\frac{k_2}{k_3}-1, k_{3} \ne 0.\tag{A11}
\label{A11}
\end{equation}
Having determined these parameters, we can easily find the eigenfunction $\Psi(s)$ expressed in terms of the hypergeometric function as
\begin{align*}
\Psi(s)&=&(-1)^nC_2\frac{\Gamma(2k_4+k_1+n)}{\Gamma(2k_4+k_1)}s^{k_4}(1-k_3s)^{k_5}\ _2F_1\left(-n, n+2(k_4+k_5)+\frac{k_2}{k_3}-1; 2k_4+k_1, k_3s\right)\nonumber\\
&=&N_ns^{k_4}(1-k_3s)^{k_5}\ _2F_1\left(-n,\ n+2(k_4+k_5)+\frac{k_2}{k_3}-1;\ 2k_4+k_1,\ k_3s\right), k_{3} \ne 0,\tag{A12}
\label{A12}
\end{align*}
where $N_n$ is the normalization constant.
Now, let us turn to the derivation of the energy eigenvalues. By using the equation (\ref{A5}) we can rewrite the $\lambda_0(s)$ and $s_0(s)$ and consequently we can calculate $\lambda_k(s)$ and $s_k(s)$. Thus, it gives
{
\begin{align*}
&&\lambda_0=\left[\frac{sk_3(2k_4+2k_5+\frac{k_2}{k_3})-(2k_4+k_1)}{s(1-k_3s)}\right]\nonumber\\
&&s_0=\left[\frac{k_3(k_4+k_5)^2+(k_4+k_5)(k_2-k_3)+\frac{A}{k_3}}{s(1-k_3s)}\right]\nonumber\\
&&\lambda_1=\left[\frac{2k_3(k_4+k5)+k_2}{s(1-k_3s)}-\frac{kk_1(1-k_3s+kk_1)}{s^2(1-k_3s)}+\frac{k_3kk_1}{s(1-k_3s)^2}+\frac{k_3(k_4+k_5)^2+(k_2-k_3)(k_4+k_5)+\frac{A}{k_3}}{s(1-k_3s)}\right]\nonumber\\
&&s_1=\left[\frac{2kk_2k_3s(1+k_4+k_5)-(1+2k_4+k1-sk_2)}{s^2k_3(1-k_3s)^2}\right]\nonumber\\
&&...etc\nonumber
\end{align*}}
where $kk_1=-2k_4-k_1+2s(k_4k_3+k_5k_3+k_2/2)$ and $kk_2=k_3(k_4+k_5)^2+k_2k_3(k_4+k_5)-k_3^2(k_5+k_4)+A$. Combining these results with the termination condition given by equation (\ref{A7}) gives
\begin{align*}
s_0\lambda_1&=&\lambda_0s_1\ \ \ \ \Rightarrow\ \ \ \ k_4+k_5=-\frac{1}{2k_3}\left(k_2-k_3-\sqrt{(k_3-k_2)^2-4A}\right),\nonumber\\
s_1\lambda_2&=&\lambda_1s_2\ \ \ \ \Rightarrow\ \ \ \ k_4+k_5=-\frac{1}{2k_3}\left(k_2+k_3-\sqrt{(k_3-k_2)^2-4A}\right),\nonumber\\ \tag{A14}
s_2\lambda_3&=&\lambda_2s_3\ \ \ \ \Rightarrow\ \ \ \ k_4+k_5=-\frac{1}{2k_3}\left(k_2+3k_3-\sqrt{(k_3-k_2)^2-4A}\right),\\
s_3\lambda_4&=&\lambda_3s_4\ \ \ \ \Rightarrow\ \ \ \ k_4+k_5=-\frac{1}{2k_3}\left(k_2+5k_3-\sqrt{(k_3-k_2)^2-4A}\right),\nonumber\\
etc.\nonumber
\label{A14}
\end{align*}
By finding the nth term of the above arithmetic progression, the formula for the eigenvalues can be obtained as
\begin{equation}
k_4+k_5=\frac{1-2n}{2}-\frac{1}{2k_3}\left(k_2-\sqrt{(k_3-k_2)^2-4A}\right)\tag{A15}
\label{A15}
\end{equation}
or more explicitly as
\begin{equation}
\left[\frac{k_4^2-k_5^2-\left[\frac{1-2n}{2}-\frac{1}{2k_3}\left(k_2-\sqrt{(k_3-k_2)^2-4A}\right)\right]^2}{2\left[\frac{1-2n}{2}-\frac{1}{2k_3}\left(k_2-\sqrt{(k_3-k_2)^2-4A}\right)\right]}\right]^2-k_5^2=0.\tag{A16}
\label{A16}
\end{equation}
\subsection*{Derivation via functional Analysis Approach}
FAA also called the traditional method by some authors has been introduced for finding bound state solutions ever since the birth of quantum mechanics. In the approach, one transform the equation of form (\ref{F1}) to a form of hypergeometric equation $_2F_1(a, b; c; s)$ via an appropriate transformation by considering the singularity points of the differential equation within the framework of the Frebenius theorem. The eigensolutions are obtained from the properties of the hypergeometric functions on the condition that the series $_2F_1(a, b; c; s)$ approaches infinity unless $a$ is a negative integer (i.e $a=-n$). This method have been employed by alot of researchers to obtain eigensolution of quantum mechanical problems in both relativistic and non relativistic case ([24] and refs. therein). Thus, solution of equation (\ref{A4B}) can now be obtained 
\begin{align*}
F(s)&=&\frac{\left[(\alpha+n)(\alpha+n-1)(\alpha+n-2)....\right]\left[(\beta+n)(\beta+n-1)(\beta+n-2).....\right]\left[\Gamma(\gamma)\right]}{\left[\Gamma(\alpha)\right]\left[\Gamma(\beta)\right]\left[(\gamma+n)(\gamma+n-1)(\gamma+n-2).....\right]}\frac{(k_3s)^n}{n!}\nonumber\\
&=&\ _2F_1(\alpha,\beta;\gamma;k_3s)=1+\sum^\infty_{n=1}\frac{(\alpha)_n(\beta_n)(k_3s)^n}{n!(\gamma)_n},\tag{A17}
\label{A17}
\end{align*}
where $\alpha$, $\beta$ and $\gamma$ are given by 
\begin{subequations}
\begin{align*}
\alpha&=&k_4+k_5-\frac{1}{2}+\frac{1}{2k_3}\left[k_2-\sqrt{(k_2-k_3)^2-4A}\right],\tag{A18a}
\label{A18A}\\
\beta&=&k_4+k_5-\frac{1}{2}+\frac{1}{2k_3}\left[k_2+\sqrt{(k_2-k_3)^2-4A}\right],\tag{A18b}
\label{A18B}\\
\gamma&=&2k_4+k_1.\tag{A18c}
\label{A18C}
\end{align*}
\end{subequations}
Considering the finiteness of the solutions, it is shown from equation (\ref{A17}) that F(s) approaches infinity unless $\alpha$ is a negative integer. Nonetheless, the wave function $\Psi(s)$ will not be finite everywhere unless we take
\begin{equation}
\alpha=k_4+k_5-\frac{1}{2}+\frac{1}{2k_3}\left[k_2-\sqrt{(k_2-k_3)^2-4A}\right]=-n,\ \ \ n=0, 1, 2, 3,...\tag{A19}
\label{A19}
\end{equation}
Hence, with equation (\ref{A19}), the expression for $\beta$ given by equation (\ref{A18B}) can be rewritten as
\begin{equation}
\beta=2(k_4+k_5)+\frac{k_2}{k_3}-1+n.\tag{A20}
\label{A20}
\end{equation}
Using equations (\ref{A19}) and (\ref{A20}), the solution of equation (\ref{A4B}) can now be expressed as
\begin{equation}
F(s)=\ _2F_1\left(-n,\ n+2(k_4+k_5)+\frac{k_2}{k_3}-1,\ 2k_4+k_1,\ k_3s\right).\tag{A21}
\label{A21}
\end{equation}
With equation (\ref{A21}), we can finally rewrite the wave function $\Psi(s)$ in equation (\ref{A3}) as
\begin{equation}
\Psi(s)=N_ns^{k_4}(1-k_3s)^{k_5}\ _2F_1\left(-n,\ n+2(k_4+k_5)+\frac{k_2}{k_3}-1;\ 2k_4+k_1,\ k_3s\right),\tag{A22}
\label{A22}
\end{equation}
where $N_n$ is the normalization constant. In addition, by using equation (\ref{A19}), we can find the following formula for energy eigenvalues:
\begin{equation}
\left[\frac{k_4^2-k_5^2-\left[\frac{1-2n}{2}-\frac{1}{2k_3}\left(k_2-\sqrt{(k_3-k_2)^2-4A}\right)\right]^2}{2\left[\frac{1-2n}{2}-\frac{1}{2k_3}\left(k_2-\sqrt{(k_3-k_2)^2-4A}\right)\right]}\right]^2-k_5^2=0, k_{3} \neq 0.\tag{A23}
\label{A23}
\end{equation} 
It should be noted that the two approaches for obtaining this formula yield the same results. Furthermore, let us discuss case where $k_3\rightarrow0$. Therefore, equation (\ref{A1}) reduces to
\begin{equation}
\Psi''(s)+\frac{(k_1-k_2s)}{s}\Psi'(s)+\frac{(As^2+Bs+C)}{s^2}\Psi(s)=0.\tag{A24}
\label{A24}
\end{equation}
Also, the proposed wave function (\ref{A3}) becomes
\begin{equation}
\lim_{k_3\rightarrow0}\Psi(s)=s^{k_4}exp(-k_5s)F(s)\tag{A25}
\label{A25}
\end{equation}
with
\begin{equation}
\lim_{k_3\rightarrow0}k_4=\frac{(1-k_1)+\sqrt{(1-k_1)^2-4C}}{2},\ \ \ \ \ \lim_{k_3\rightarrow0}k_5=-\frac{k_2}{2}+\sqrt{\left(\frac{k_2}{2}\right)^2-A}.\tag{A26}
\label{A26}
\end{equation}
By following the same procedure, the eigenvalues and the corresponding wave function can be obtained as
\begin{equation}
\left[\frac{B-k_4k_2-nk_2}{2k_4+k_1+2n}\right]^2-k_5^2=0,\tag{A27}
\label{A27}
\end{equation}
and
\begin{equation}
\Psi(s)=N_ns^{k_4}exp(-k_5s)\ _1F_1\left(-n,2k_4+k_1, (2k_5+k_2)s\right),\tag{A28}
\label{A28}
\end{equation}
respectively. $N_n$ is the normalization constant.
\end{changemargin}
\end{document}